\documentclass[preprint,12pt]{elsarticle}

\usepackage{amsmath,amssymb,array,calc,rotating,epsfig,psfrag,empheq}

\title{On the scaling of the distribution of daily price fluctuations in Mexican financial market index}

\author{L\'ester Alfonso$^a$,Ricardo Mansilla$^b$, C\'esar A. Terrero-Escalante$^c$\corref{cor1}}
\cortext[cor1]{Corresponding author: cterrero@ucol.mx}
\address{$^a$ Universidad Aut\'onoma de la Ciudad de M\'exico,\\ C.P. 09790, M\'exico D.F.,  M\'exico\\
$^b$ Centro de Investigaciones Interdisciplinarias en Ciencias y Humanidades,\\ 
Universidad Nacional Aut\'onoma de M\'exico,\\ 
Ciudad Universitaria, C.P. 04510, M\'exico~D.F.,~M\'exico.\\
$^c$Facultad de Ciencias, Universidad de Colima,\\
Bernal D\'\i az del Castillo 340, Col. Villas San Sebasti\'an,\\ 
C.P. 28045, Colima, Colima, M\'exico.\\
}

\journal{Physica A}
\date{\today}

\begin{document}

\begin{frontmatter}

\begin{abstract}
In this paper, a statistical analysis of log-return fluctuations of the IPC, the Mexican Stock Market Index is presented. 
A sample of daily data covering the period from $04/09/2000-04/09/2010$ was analyzed, and fitted to different distributions. 
Tests of the goodness of fit were performed in order to quantitatively asses the quality of the estimation.
Special attention was paid to the impact of the size of the sample on the estimated decay of the distributions tail. 
In this study a forceful rejection of normality was obtained.
On the other hand, the null hypothesis that the log-fluctuations are fitted to a $\alpha$-stable L\'evy distribution 
cannot be rejected at $5\%$ significance level.
\end{abstract}

\begin{keyword}
stock markets \sep return fluctuations distribution \sep tail behavior \sep $\alpha$-stable processes

\MSC 91B82 \sep 60E07 \sep 62D05

\end{keyword}

\end{frontmatter}

\section{Introduction}
\label{sec:Intro}

Today is widely acknowledged that for the proper management of assets and prices (and the related investment risks) it is required the proper modeling of the return distribution of financial assets. 
For instance, the answer to whether it is possible to beat the market except by chance depends on whether stock market prices display long memory and how probable are very large price fluctuations. 
The crucial difficulty is, however, that the financial market is a very complex system; 
it has a large number of non-linearly interacting internal elements, and is highly sensible to the action of external forces. 
Even more, the real challenge here is that the number of the system constituents, and the details of their interactions and of the external factors acting upon it are actually barely known. 

Physicists have a long tradition of dealing with similar systems. 
The statistical description of systems of many particles was developed in parallel with the statistical analysis of market dynamics \cite{pareto,bachelier,levy}. 
For instance, taking into account the wide applicability of the Central Limit Theorem, Bachelier 
assumed that the return over a given time scale is the consequence of many independent ``microscopic" events, 
which then lead to a normal distribution of returns. Thus, he modeled their dynamics as an uncorrelated random walk with independent, identically Gaussian distributed random variables, 
i.e., as a Brownian motion \cite{bachelier}. 
Since then, the Gaussian assumption for the distribution of returns has been frequently used in mathematical finance and is one of the key assumptions behind the classic Black-Scholes option pricing formula \cite{black-scholes}, 
which is based on a Wiener process in the continuous-time setting or on appropriate discrete-time versions such as binomial trees. 

\begin{figure}[!ht]
\centering
\epsfig{figure=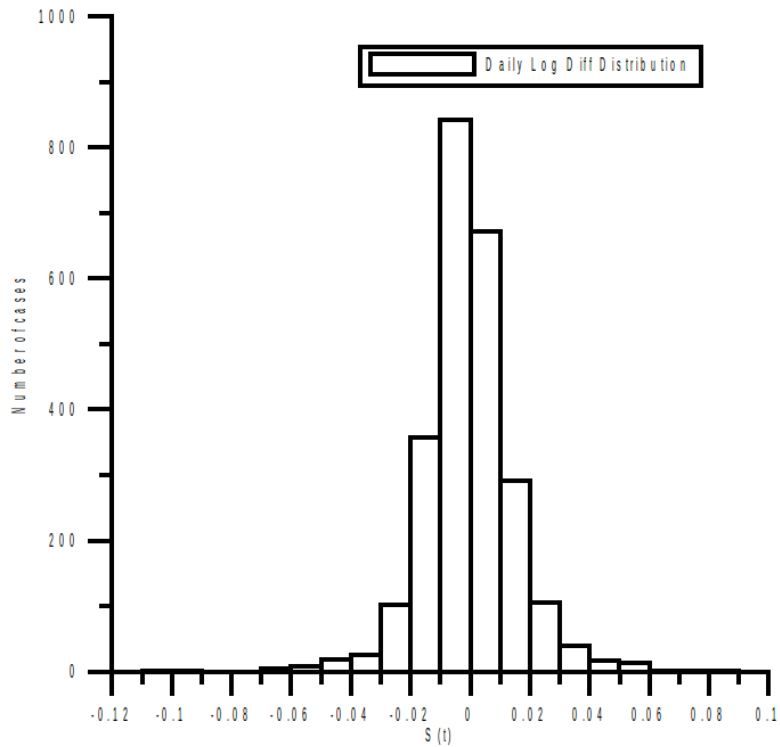, width=0.75\linewidth, angle=0}
\caption{Histogram for daily logarithm differences of the Mexican IPC index from April 9th, 2000 to April 9th, 2010.}
\label{fig:histo}
\end{figure}

Thought the simplifications the normal distribution provides in analytical calculation are very valuable, empirical studies \cite{mandelbrot,mantegna-stanley,gopikrishnan,cont} 
show that the distribution of returns has a tail heavier than that of a Gaussian. 
To illustrate this fact, we show in fig.\ref{fig:histo}
the histogram for daily logarithm differences of the Mexican IPC index from April 9th, 2000 to April 9th, 2010.
Clearly, large events are very frequent in the data, a fact largely underestimated by a Gaussian process and of upmost importance for financial management. 
It is remarkable that such a feature is present in quite different markets \cite{mandelbrot,mantegna,gopikrishnan}.

One of the skills of physicists is the search for universal laws, i.e., common features in most particular realizations of a general class of phenomena. 
This can be done even if the ``microscopic" details distinguishing each case are not fully understood. 
Heavy tailed distributions are commonly described by a power law (at least in a range of scales), which in turn implies scale invariance, a distinct signature of fractals. 
Fractals have been shown to be a common geometrical pattern in many natural systems. 
Finally, many of these systems may be in a state of self-organized criticality, a paradigm that would explain how organization arises in complex systems and make them more predictable. 
This could be also the case for other dynamical systems outside the realm of natural sciences.  
In his pioneering analysis of cotton prices, Mandelbrot \cite{mandelbrot}
(the founder of fractal geometry) observed that in addition to being non-Gaussian, the process of returns shows another interesting property: 
time scaling, that is, the distributions of returns for various choices of $t$, ranging from one day up to one month have similar functional forms.  
As it was already mentioned, observed stock market prices are assumed to be the sum of many small terms, hence a statistical model to describe them must be such that the sum of two independent random variables having the given distribution (with a parameter $\alpha$ describing the decay of the tail) yields again the same kind of distribution (with the same value of $\alpha$). 
Motivated by these empirical findings and reasoning, Mandelbrot proposed that the returns can be modeled as a kind of stable process introduced by L\'evy in 1925\cite{levy}. 
L\'evy stable distributions are attractive because they are supported by the generalized Central Limit Theorem. 
The theorem states that stable laws are the only possible limit distributions for properly normalized and centered sums of independent, identically distributed random variables. 

An issue here is whether the underlying distributions are actually stable. 
Stability only holds for $\alpha \in (0, 2]$ and some authors have found that the tails of some financial time series have to be modeled with $\alpha > 2$ 
\cite{gopikrishnan,pagan,coronel}. 
Conclusive results on the distribution of returns are difficult to obtain, and require a large amount of data to study the rare events that give rise to the fat tails \cite{weron}. 
Another issue is that according with Cont \cite{cont}, in order for a parametric distributional model to reproduce the properties of the empirical distribution it must have at least four parameters: 
a location parameter, a scale parameter, a parameter describing the decay of the tails and an asymmetry parameter. 
And we know of several other heavy-tailed alternative distributions (such as student's t, hyperbolic or normal inverse Gaussian) which fullfil this condition. 
Therefore, in order to grasp the universal laws behind markets dynamics, it is important to keep accumulating empirical facts about the statistics of different financial indices around the world. 

The aim of this paper is to provide a rigourous statistical analysis of log-return daily fluctuations of the IPC, the leading Mexican Stock Market Index. 
After considering the Gaussian, normal inverse Gaussian and L\'evy distributions, we found that this last provides the best fit and we show that the corresponding $\alpha$ 
is indeed in the range for a stable L\'evy distribution.

The paper is organized as follows. In the next section we review the basic properties of L\'evy-stable distributions.
In section \ref{sec:stats} we proceed to introduce and analize the IPC data. 
In section \ref{sec:comparison} we discuss the similarities and disagreements between our results and results in previous works.
Finally we present our conclusions in section \ref{sec:conclusions}.

\section{Stable distributions}
\label{sec:stable}

L\'evy-stable distributions were introduced by Paul L\'evy \cite{levy} during his investigations of the behavior of sums of independent random variables. 
A L\'evy skew alpha-stable distribution is specified by scale $\gamma$, exponent $\alpha$, skewness parameter $\beta$ and a location parameter $\mu$. 
Since the analytical form of the Levy stable distribution is known only for a few cases, 
they are generally specified by their characteristic function. 
The most popular parameterization is defined by Samorodnitsky, G. and Taqqu, \cite{taqqu} with the characteristic function:
\begin{equation}
  \phi(t)=\begin{cases}
    \exp{\left(-\gamma |t| \left[1+i\beta\frac2{\pi }\rm{sign}(t) \ln(|t|) + i\mu t\right]
\right)}, & \text{if $\alpha=1$}.\\
    \exp{\left(-\gamma^\alpha |t|^\alpha \left[1-i\beta\tan\left(\frac{\pi \alpha}2\right) {\rm sign}(t) + i\mu t\right]
\right)}, & \text{otherwise}.
  \end{cases}
\end{equation}
where $sign(t)$ stands for the sign of $t$. 
Then, the probability density function is calculated from with the inverse fourier transform in the form: 
\begin{equation}
f(x;\alpha,\beta,\gamma,\mu)=\frac1{2\pi}\int^{+\infty}_{-\infty}\phi(t){\rm e}^{-itx} dt  \, .
\end{equation}	

L\'evy distributions are characterized by the property of being stable under convolution, 
i.e, the sum of two independent and identically L\'evy-distributed random variables, is also L\'evy distributed with the same stability index $\alpha$. 	
The stability parameter $\alpha$ lies in the interval $(0, 2]$. 
Small $\alpha$ represents a sharp peak but heavy tails which asymptotically decay as power laws with exponent $-(\alpha+1)$. For the normal distribution $\alpha=2$. 
For the symmetric distributions (like the normal distribution), the skewness parameter $\beta=0$. The skewness parameter must lie in the range $[-1, 1]$. 
When $\beta=+1,-1$, one tail vanishes completely. 
The parameter $\gamma$ lies in the interval $(0, \infty)$, 
while the location parameter $\mu$ is in $(-\infty,+\infty)$.

\begin{figure}[!ht]
\centering
\epsfig{figure=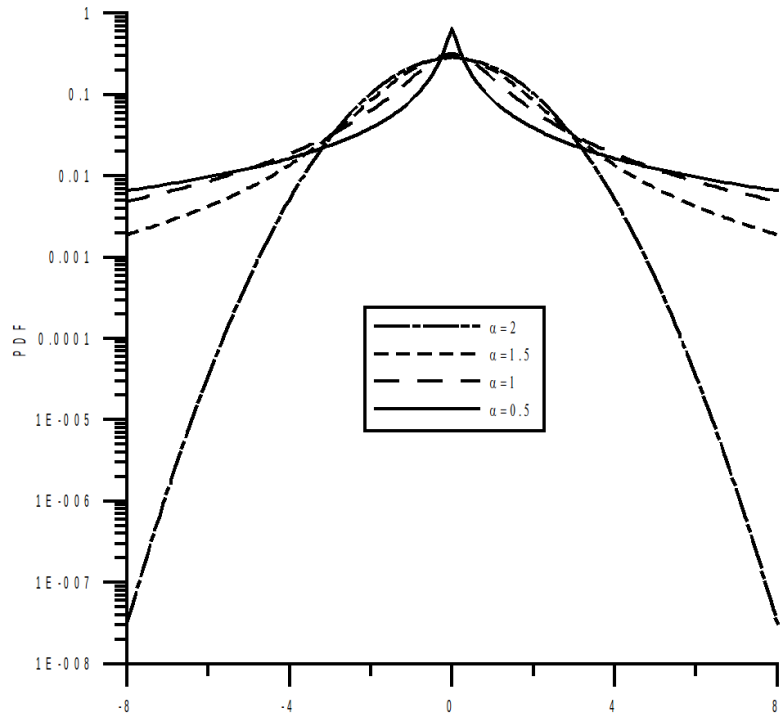, width=0.75\linewidth, angle=0}
\caption{A semilog plot of symmetric ($\beta=\mu=0$) L\'evy-stable probability distribution function for $\alpha$=2, 1.5, 1 and 0.5. 
Only the normal distribution has exponential tails.}
\label{fig:levypdf}
\end{figure}

The asymptotic behaviour of the L\'evy distributions is described by the expression
\begin{equation}
f(x;\alpha)\approx |x|^{-1-\alpha} \, .
\end{equation}	                                                                                                                   
Hence, the variance of the Levy stable distributions is infinite for all $\alpha<2$. 
The dependence on $\alpha$ is illustrated in the semilog plots in figure \ref{fig:levypdf}. 

In our computations the stable library developed by Nolan was used.

\section{Statistical analysis of data}
\label{sec:stats}

\begin{figure}[!ht]
\centering
\epsfig{figure=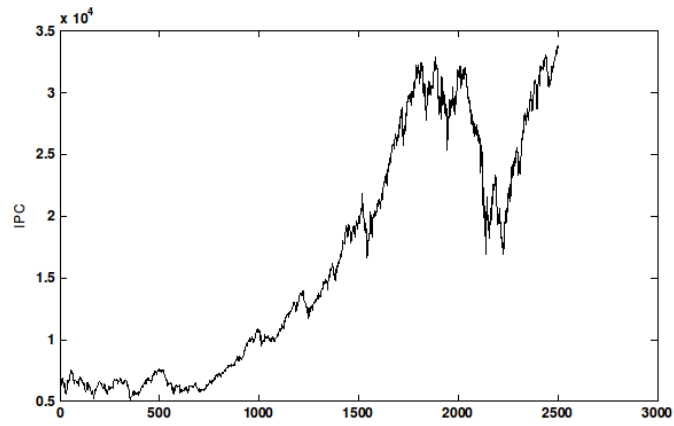, width=0.75\linewidth, angle=0}
\caption{IPC evolution for the 10 year period 04/9/2000-04/9/2010.}
\label{fig:series}
\end{figure}

Our IPC data set covers the period from April, 2000 to April, 2010, and comprises 2502 returns. 
The index was label as $Y(t)$ and the data is written as the successive differences of the natural logarithm of the returns,
\begin{equation}
S(t) = \ln Y(t+\Delta t) - \ln Y(t) \, .
\end{equation}                                            
The daily closure values of the IPC were used, so $\Delta t=1$ day. 
Fig.~\ref{fig:series} 
displays the index value as a function of trading day and the corresponding histogram for $S(t)$ is presented in figure \ref{fig:histo}. 
In table \ref{tab:stats}
\begin{table}[ht]
\centering
\begin{tabular}{|l|c|c|c|c|}
\hline
Index & Mean & Standard Deviation & Skewness & Kurtosis\\
\hline
IPC & $-0.609\times10^{-3}$ & $0.151\times10^{-1}$ & $-0.395\times10^{-1}$ & $7.23$\\
\hline
\end{tabular}
\caption{Mean variance, skewness and kurtosis of the IPC log-return time series.}
\label{tab:stats}
\end{table}
the mean, standard deviation, skewness and kurtosis are reported for the IPC log-return time series. 
The high value of the kurtosis ($7.23$) shows that the density functions of the time series is more peaked than in Gaussian distributions.

\subsection{Gaussian fit}
\begin{figure}[!ht]
\centering
\epsfig{figure=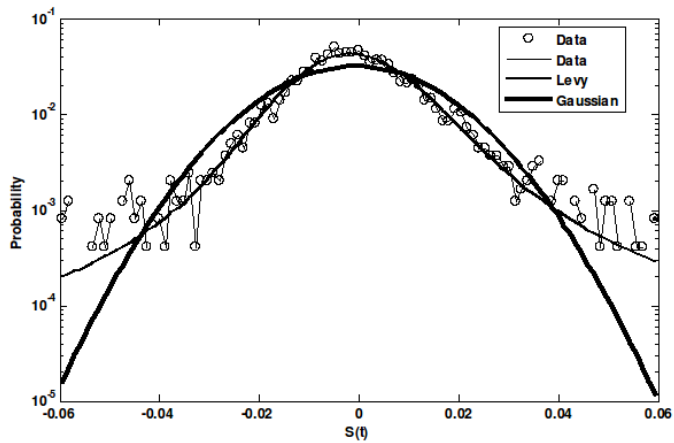, width=0.75\linewidth, angle=0}
\caption{Fits of daily logarithm differences to the Gaussian distributions and $\alpha$-stable Levy.}
\label{fig:fits}
\end{figure}

The method of maximum likelihood was used to determine the parameters of the Gaussian distribution that fits the experimental data.
The results are shown in table \ref{tab:params},
\begin{table}[ht]
\centering
\begin{tabular}{|l|c|c|c|c|c|}
\hline
Parameters & $\alpha$ & $\beta$ & $\gamma$ & $\mu$ & $\sigma$\\
\hline
Gaussian fit &  &  &  & $-6.09\times10^{-4}$ & $0.0151$\\
\hline
$\alpha$-stable fit & $1.64$ & $0.219$ & $0.00815$ & $-0.000186$ & \\
\hline
NIG fit & $55.43$ & $-0.2990$ & $0.01254$ & $-0.000541$ & \\
\hline
\end{tabular}
\caption{Estimated parameters of the normal, $\alpha$-stable and NIG distributions.}
\label{tab:params}
\end{table}
where $\mu$ is the mean and $\sigma$ the standard deviation. 
In figure \ref{fig:fits}
the empirical PDF for the daily data is shown along with the fitted Gaussian pdf. 
As it can be seen, 
it is not a good fit, especially in the tails of the distribution.

Now we will verify the hypothese, $H_0$, that the sample comes from a Gaussian distribution with the above estimated parameters. 
Using the $n=2502$ sample size, the hypothese is examined by three criteria: 
The chi-square goodness of fit test, the Anderson-Darling test and the Kolmogorov-Smirnov (K-S) method. 

For the Pearson test, the experimental statistics is $\bar{\chi}^2=161.9$ with 2 degrees of freedom, 
then the null hypothesis $H_0$ for this case (sample comes from a normal distribution) can be rejected at $5\%$ significance level.
In turn, the Anderson-Darling test yield an experimental statistics of $26.6115$ and a zero p-value, clearly rejecting 
$H_0$ at $5\%$ significance level.

The experimental statistics for the K-S test can be obtained by arranging the data in ascending order $(x_1, x_2,\cdots, x_n)$, 
and deriving the maximum difference between the rank statistics $(i-1)/n$ and the theoretically calculated cdf :
\begin{equation}
D=\max\left(\max\left[\left|F(x_i)-\frac{i-1}n\right|,\left|\frac{i}n - F(x_i)\right|\right]\right) \, .
\label{eq:kss}
\end{equation}                                
Given that the parameters of the fitting distributions were estimated from the observed data, limiting values provided by the K-S criteria cannot be used. 
In this case, approximate p-values can be obtained via Monte Carlo simulations \cite{borak}.
First, the parameter vector is estimated for a given sample of size $n$, $\hat{\theta}=(\hat{\alpha},\hat{\beta},\hat{\gamma},\hat{\mu})$ being the result, 
and the test statistics is calculated assuming that the sample is distributed according to $F(x;\hat{\theta})$, returning a value of $D$. 
Next, a sample of size $n$ $F(x;\hat{\theta})$ variates is generated and the parameter vector $\hat{\theta}_1$ is estimated. 
The test statistics is again calculated assuming that the sample is distributed according to $F(x;\hat{\theta}_1)$. 

Such a calculation was made for $n=2502$ (our sample size) 1000 times, and the distribution pattern of $D_i$ was derived. 
Then, $5\%$ percent point from the greater side was taken as the estimated limiting values. 
The estimate of p-value is calculated as the relative number of occasions in which the test statistics is at least as large as $D$. 
The test statistics and the corresponding estimated p-values are shown in table \ref{tab:pvalues}. 
\begin{table}[ht]
\centering
\begin{tabular}{|l|c|c|c|c|}
\hline
Case & K-S Test   & K-S (estimated) & p-value & Reject $H_0$?\\
     & Statistics & limiting values &  & \\
     &  & for $\alpha=0.05$ &  & \\
\hline
Gaussian  & $0.0651$  & $0.0184$ & $0.000$ & Yes\\
\hline
$\alpha$-stable L\'evy & $0.0165$ & $0.0209$ & $0.133$ & No\\
\hline
NIG & $0.0272$ & $0.0207$ & $0.01$ & Yes\\
\hline
\end{tabular}
\caption{Test statistics, K-S criterion limiting values and the corresponding p-values based on 1000 simulated samples for the normal, $\alpha$-stable Levy and NIG fits.}
\label{tab:pvalues}
\end{table}
At $5\%$ significance, the experimental statistics can be compared with the $95\%$ limiting values obtained from the Monte Carlo simulations. 
The values of the test statistics ($0.0651$) yield p-values smaller than $0.05$ forcing us, again, to reject the null hypothesis.
Summarizing, the three tests we have used reject the Gaussian distribution as the underlying statistics.

\subsection{Alpha-stable fit}

Using the method of maximum likelihood, the parameters of the Levy skew $\alpha$-stable distribution that fits the experimental data were found.
Their values are presented in table \ref{tab:params}, 
where $\alpha=1.64$ is the index, $\beta=0.219$ the skewness parameter, $\gamma=0.00815$ a scale factor, and $\mu= -0.000186$ a location parameter.
The plot of the fitted Levy pdf presented in figure \ref{fig:fits} 
shows good agreement with the empirical distribution. 
Now we will verify the hypothese, $H_0$, that the sample comes from a L\'evy distribution with the above estimated parameters. 
Since the Anderson-Darling p-values for this distribution are not available, we will use
the chi-square goodness of fit test, and the K-S method. 

The Pearson statistics is $\bar{\chi}^2=16.27$ with 15 degrees of freedom. 
The probability $\chi^2 > \bar{\chi}^2$ was found to be $0.636$. 
Therefore, at $5\%$ significance level, the null hypothesis $H_0$ 
(sample comes from $\alpha$-stable L\'evy distribution) 
cannot be rejected. 

The KS test statistics and the corresponding estimated p-values for the $\alpha$-stable Levy distribution 
are shown in table \ref{tab:pvalues}. 
Again, at $5\%$ significance, the experimental statistics can be compared with the $95\%$ limiting values obtained from the Monte Carlo simulations. 
For the L\'evy fit, the experimental statistics is $D=0.0165$, which is smaller than the limiting value $D_{1-\alpha} =0.0209$. 
Therefore, the null hypothesis $H_0$ (sample comes from $\alpha$-stable Levy distribution) cannot be rejected at the $5\%$ significance level. 
The corresponding p-value obtained from the Monte Carlo simulation was found equal to $0.133$. 

It is worthy to note here that the statistics and p-values obtained via Monte Carlo simulations are quite different from those commonly used as reference values
is standarized test for normality. This supports that the correct procedure is to estimate these values in a case-by-case basis.

\subsection{Normal-inverse Gaussian fit}

The hyperbolic distribution also provides the possibility of modeling fat tails.
The normal-inverse Gaussian (NIG) distributions were introduced as a subclass of the generalized hyperbolic laws (GH), 
and were developed \cite{barndorff} for modeling the grain size distribution of windblown sand. 
Empirical studies \cite{barndorff2} suggest a good fit of the NIG law to financial data. 
In fact, in Ref.\cite{nunez} it was found that the NIG distribution fits nicely the IPC data and other Mexican finantial indices. 
Therefore, we will also fit the data to a NIG and compare our results with the $\alpha$-stable and Gaussian fits. 

The GH distribution is specified by scale $\gamma$, exponent $\alpha$, skewness parameter $\beta$, a location parameter $\mu$, and a parameter $\lambda$ 
that characterizes the subclasses of the GH distribution. 
Its probability density function is given by:
\begin{equation}
f_{HG}(x;\alpha,\beta,\delta, \lambda,\mu)= \kappa\left[\delta^2
+(x-\mu)^2\right]^{\frac12(\lambda-\frac12)} K_{\lambda-\frac12}\left(\alpha\sqrt{\delta^2+(x-\mu)^2}\right) {\rm e}^{\beta(x-\mu)}\, ,
\end{equation}	 
where:
\begin{equation}
\kappa=\frac{(\alpha^2-\beta^2)^\frac{\lambda}2}{\sqrt{2\pi}\alpha^{\lambda-\frac12}\delta^{\lambda}K_0(\delta\sqrt{\alpha^2-\beta^2})}\, .
\end{equation}	 
The NIG distribution corresponds to $\lambda=-1/2$ 
and is able to model symmetric and asymmetric distributions with possibly long tails in both directions. 
It has the probability density function:
\begin{equation}
f_{NIG}(x;\alpha,\beta,\delta,\mu)= \frac{\alpha\delta}\pi {\rm e}^{\delta\sqrt{\alpha^2-\beta^2}+\beta(x-\mu)} \frac{K_{-1}(\alpha\sqrt{\delta^2+(x-\mu)^2)}}{\sqrt{\delta^2+(x-\mu)^2}}
 \, .
\end{equation}	 
Our fit to the IPC data yields $\alpha=55.43$, $\beta=-0.2990$, $\delta=0.01254$ and $\mu=-0.000541$
(see table \ref{tab:params}). 
The Kolmogorov-Smirnov experimental statistics (\ref{eq:kss}) was found to be $D=0.0272$, 
which is larger than the $95\%$ limiting value ($D_{1-0.05} =0.0207$) obtained from a $1000$ realization Monte Carlo simulation. 
For this case, the values of the test statistics ($0.01$) yield p-values smaller than $0.05$ forcing us to reject the null hypothesis for a $5\%$ significance level. 
It must be mentioned, nevertheless, 
that the null hypothesis (sample comes from a NIG distribution) could not be rejected at the $1\%$ significance level. 

\section{Comparison with previous studies and discussion}
\label{sec:comparison}

In \cite{coronel} a sample of IPC covering the period 13-year period $04/19/1990 -- 08/21/2003$ was analyzed. 
The authors claimed that the cumulative distribution function for extreme variations can be described by a Pareto-L\'evy model 
with shape parameters $\alpha=3.634\pm0.272$ and $\alpha=3.540\pm0.278$ for its positive and negative tail respectively. 
As a consequence they concluded that the process that governs the time series is well outside the L\'evy regime. 
However, as was reported in [10], a value of the tail exponent of about $3$ may very well indicate a L\'evy-stable distribution with $\alpha$ about $1.8$. 
Hence, the results obtained in \cite{coronel} could be a consequence of working with small samples.
Following \cite{weron}, we generated samples of size $N=10^3$, $2502$, $10^4$, $10^5$, $10^6$ and $10^7$ 
of L\'evy-stable distributed random variables with parameters $\alpha=1.7$ 
(which is close to $\alpha=1.64$ obtained for our IPC data), $\beta=0$, $\gamma=1$ and $\mu=0$. 
Then, the tail $\alpha$ index was estimated \cite{weron} with the method developed in \cite{mantegna} 
that combines maximum-likelihood fitting methods with goodness of fit tests based on the Kolmogorov-Smirnov statistics and likelihood ratios. 
The results of the simulations are displayed in table \ref{tab:longrun}.
\begin{table}[ht]
\centering
\begin{tabular}{|c|c|}
\hline
Number of samples & $\alpha$\\
\hline
$10^3$ &  $2.166$\\
\hline
$2502$ &  $2.137$\\
\hline
$10^4$ &  $1.853$\\
\hline
$10^5$ &  $1.771$\\
\hline
$10^6$ &  $1.613$\\
\hline
$10^7$ &  $1.714$\\
\hline
\end{tabular}
\caption{Estimated $\alpha$ as funtion of the samples size.}
\label{tab:longrun}
\end{table}
As can be observed in table  \ref{tab:longrun}, for $N=2502$ a value of $\alpha=2.137$ was obtained which is also outside the Levy regime. 
These simulations suggest that high frequency data is needed in order to obtain reliable results while estimating the tail index[12].

The results obtained in section 3.1 are in general in agreement with previous results obtained in \cite{nunez}, where a better fit than the Gaussian was obtained for the NIG distribution. 
In that paper the data set under analysis covered the period January, 1993-January 2003, and comprises 2505 returns. 
The KS goodness of fit analysis was performed at the $1\%$ significance level.
A clear rejection of normality was obtained by those authors. 
At this significance level their fit of the data using a NIG was not rejected.
Nevertheless, our results clearly show that the $\alpha$-stable distribution provides a better fit than the NIG distribution for daily fluctuations of the IPC data.

\section{Conclusions}
\label{sec:conclusions}

The three tests we have used forceful rejected that the data of the Mexican financial index, IPC, is normally distributed.
Likewise, at the $5\%$ significance level, the corresponding tests force us to reject the null hypothesis of the data being distributed as a normal inverse Gaussian.
On the other hand, the null hypothesis that our sample comes from $\alpha$-stable Levy distribution cannot be rejected at the $5\%$ significance level. 
Our analysis of the impact of the sample size on the estimation of the tail index
confirms that high frequency data is needed in order to determine whether or not a given distribution is stable. 
Of the three distributions we have considered, the $\alpha$-stable distribution provides the best fit for daily fluctuations of the IPC data. 

\section{Acknowledgments}

This research was supported by the Sistema Nacional de Investigadores (M\'exico). 
The work of CAT-E was also partially funded by PROMEP (M\'exico) under grant PROMEP/103.5/10/4948.

\end{document}